# Multifrequency multi-qubit entanglement based on plasmonic hot spots

Jun Ren, Tong Wu and Xiangdong Zhang[*]

School of Physics, Beijing Institute of Technology, 100081, Beijing, China

[*] *Corresponding author, E-Mail address:zhangxd@bit.edu.cn*

## Abstract

The theoretical method to study strong coupling between an ensemble of quantum emitters (QEs) and surface plasmons excited by the nanoparticle cluster has been presented by using a rigorous first-principles electromagnetic Green's tensor technique. We have demonstrated that multi-qubit entanglement for two-level QEs can be produced at different frequencies simultaneously, when they locate in hot spots of metallic nanoparticle clusters. The duration of quantum beats for such an entanglement can reach two orders longer than that for the entanglement in a photonic cavity. The phenomenon originates from collective coupling resonance excitation of the cluster. At the frequency of single scattering resonance, the entanglement cannot be produced although the single QE spontaneous decay rate is very big.

# I. INTRODUCTION

Quantum entanglement plays a key role in quantum information processing such as quantum teleportation[1], quantum cryptographic[2], quantum dense coding[3] and parallel computing[4]. Entanglement between two particles has been well understood and used in different physical systems for a variety of tasks.[5-9] In contrast to the bipartite case, the deep understanding on the multipartite entanglement is still in its infancy.[9] The generation of genuine multipartite entanglement has been demonstrated only in a few physical systems such as ion traps[10-13], photon systems[14-16]. Effective creation of genuine multipartite entangled states in various physical systems is still a challenge.

On the other hand, the interaction between quantum emitters (QEs) and surface plasmon polaritons (SPPs) has attracted great interest recently.[3-5] Resonant excitation of SPPs by external optical fields allows metallic nanostructures to concentrate electromagnetic (EM) field in subwavelength volumes, resulting in enormous field enhancement especially at the so-called hotspots, e.g. the nanogaps between adjacent metallic nanoparticles.[17-19] When QEs are put in the hotspots, it has been shown that the strong coupling between them and SPPs can appear.[20,21] Up to now, many interesting theoretical and experimental works on the strong coupling between SPPs and QEs have been widely carried out.[22-30] Recently, quantum entanglement generation between two separated QDs mediated by a plasmonic waveguide has been reported.[31-36] However, these investigations only focus on the bipartite entanglement.

Motivated by these recent developments in plasmonics and quantum information science, in this work we study the generation of multipartite entanglement due to the strong interaction between the nanoparticle cluster and QEs in the hotspots. The rest of this paper is arranged as follows. In Sec. II, we present general description on the theory. Two-qubit, three-qubit and four-qubit entanglements are discussed in Sec. III, IV and V, respectively. A summary is given in Sec. VI.

# II. GENERAL DESCRIPTION ON THEORY

We consider N two-level QEs located in the cluster of nanoparticles. Let us further assume that the QEs are sufficiently far from each other, so that interatomic Coulomb interactions can be ignored. Under the electric-dipole and rotating wave approximations, the Hamiltonian of the system can be expressed as[24,37,38]

$$\hat{H} = \int d^3\vec{r} \int_0^\infty d\omega \hbar\omega \hat{\vec{f}}^\dagger(\vec{r},\omega)\hat{\vec{f}}(\vec{r},\omega) + \sum_A \frac{1}{2}\hbar\omega_A \hat{\sigma}_{Az} - \sum_A [\hat{\sigma}_A^\dagger \hat{\vec{E}}^{(+)}(\vec{r}_A)\vec{d}_A + H.c.],  \quad (1)$$

where $\hat{\vec{f}}^\dagger(\vec{r},\omega)$ and $\hat{\vec{f}}(\vec{r},\omega)$ are referred to as the creation and annihilation operators of the radiation field, respectively. The $\omega_A$ and $\vec{r}_A$ are the transition frequency and position vector of the Ath QE, $|e\rangle$ and $|g\rangle$ respectively represent its excited and ground states, $\hat{\sigma}_A^\dagger = |e\rangle_{AA}\langle g|$, $\hat{\sigma}_A = |g\rangle_{AA}\langle e|$, and $\hat{\sigma}_{Az} = |e\rangle_{AA}\langle e| - |g\rangle_{AA}\langle g|$ are Pauli operators, $\vec{d}_A$ is the dipole moment. The electric field operator in Eq. (1) is[24, 37, 38]

$$\hat{\vec{E}}^{(+)}(\vec{r}) = i\sqrt{\frac{\hbar}{\pi\varepsilon_0}}\int_0^\infty d\omega \frac{\omega^2}{c^2}\int d^3\vec{r}'\sqrt{\varepsilon_I(\vec{r}',\omega)}\overline{\overline{G}}(\vec{r},\vec{r}',\omega)\hat{\vec{f}}(\vec{r}',\omega), \quad (2)$$

where $\overline{\overline{G}}(\vec{r},\vec{r}',\omega)$ is the classical Green tensor of the system, which can be obtained exactly from the T-matrix method (see supplementary material). Here $\varepsilon(\vec{r}',\omega) = \varepsilon_R(\vec{r}',\omega) + i\varepsilon_I(\vec{r}',\omega)$ represents the complex permittivity. For a single-quantum excitation, the system wave function at time t can be written as[37, 38]

$$|\psi(t)\rangle = \sum_A C_A(t)e^{-i(\omega_A-\bar{\omega})t}|A\rangle|\{0\}\rangle \\ + \int d^3\vec{r}\int_0^\infty d\omega[C_{Li}(\vec{r},\omega,t)e^{-i(\omega_A-\bar{\omega})t}|L\rangle|\{1_i(\vec{r},\omega)\}\rangle], \quad (3)$$

where $\bar{\omega} = \frac{1}{2}\sum_A \omega_A$, $|A\rangle$ represents the upper state of Ath QE and all the other QEs are in the lower state, $|\{0\}\rangle$ is the vacuum state of the medium-assisted field, $|L\rangle$ is the lower state of QEs and here $|\{1_i(\vec{r},\omega)\}\rangle = \hat{f}_i^\dagger(\vec{r},\omega)|\{0\}\rangle$ is the state of the field which is excited in a single-quantum Fock state, $C_A(t)$ and $C_{Li}(\vec{r},\omega,t)$ are the probability amplitudes of the states $|A\rangle|\{0\}\rangle$ and $|L\rangle|\{1_i(\vec{r},\omega)\}\rangle$.

According to the method described in Ref. 37, if the new states $|i\rangle = \sum_{A'}^N x_{A'}|A'\rangle$ $(i=1,N)$ are introduced, at the same time the interference term of decay rate of any two QEs $\Gamma_{AA'} = \frac{2k_A^2}{\hbar\varepsilon_0}\vec{d}_A \text{Im}\overline{\overline{G}}(\vec{r}_A,\vec{r}_{A'},\omega_A)\vec{d}_{A'}$ and the spontaneous decay rate of single QE $\Gamma = \Gamma_{AA}$ can be obtained, the Schrodinger equation based on Eqs. (1) and (3) for the present system can be solved, namely the probability amplitudes $C_i = \sum_{A'}^N x_{A'}C_{A'}$ can be obtained. Then, if one of these new states $|i\rangle$ is in strong coupling and other states are in weak coupling, the corresponding spontaneous decay rate to $|i\rangle$, $\Gamma_i$, is very large and the decay rates for other states are nearly equal to zero. Then, the state $|i\rangle$ is often called superradiant state and other states are subradiant states. For these subradiant states, the probabilities vanish under the initial conditions $C_i(0) = 0$ $(i = 1, N)$. Thus, after tracing out the medium-assisted field, the density operator of N qubits can be written as

$$\hat{\rho} = |C_i(t)|^2 |i\rangle\langle i| + [1 - |C_i(t)|^2] |L\rangle\langle L|. \tag{4}$$

According to Ref. 39, the genuine multipartite entanglement is defined as

$$E_G^{(2)} = \frac{2}{N(N-1)} \sum_{l=1}^{N-1} (N-l) G(2,l), \tag{5}$$

where the function $G(2,l)$ is defined as

$$G(2,l) \equiv \frac{4}{3}[1 - \frac{1}{N-l} \sum_{j=1}^{N-l} \mathrm{Tr}(\rho_{j,j+l}^2)] \tag{6}$$

and $\rho_{j,j+l}$ is the reduced density matrix of qubits $j$ and $j+l$, which is obtained by tracing out the other $N-2$ qubits. Based on Eqs. (4)-(6), we can calculate the genuine multi-qubit entanglement once $C_i(t)$ has been obtained.

## III. TWO-QUBIT ENTANGLEMENT

We consider two two-level QEs located at the hotspots of linear nanosphere trimer as shown in the inset of Fig. 1(a), the radii of three spheres are taken as R and the separation distances (gaps) between them are marked by d, two QEs A and B are inserted in the gaps, and the orientations of their electric-dipole moments $\vec{d}_A$ and $\vec{d}_B$ are both along the axis of the trimer. The new states $|i\rangle$ $(i = 1,2)$ are taken as $|1\rangle = \frac{1}{\sqrt{2}}(|A\rangle + |B\rangle)$ and $|2\rangle = \frac{1}{\sqrt{2}}(|A\rangle - |B\rangle)$, their corresponding probability amplitudes are $C_1(t) = \frac{1}{\sqrt{2}}[C_A(t) + C_B(t)]$ and $C_2(t) = \frac{1}{\sqrt{2}}[C_A(t) - C_B(t)]$, and the decay rates are $\Gamma_1 = \Gamma + \Gamma_{AB}$ and $\Gamma_2 = \Gamma - \Gamma_{AB}$. The excitation can initially resides in one of the QEs or the medium-assisted field. If the excitation resides in the medium-assisted field initially, that is $C_A(0) = C_B(0) = 0$ and $C_1(0) = C_2(0) = 0$, which can be realized by coupling the field first to another excited QE E with the probability amplitude $C_E(t)$ in a time interval $\Delta t$. If the QE E locates at the same position with the QE A, and consider A, B and E obey the same decay law, from the Schrodinger equation, the corresponding integro-differential equation of $C_1(t)$ is[38]

$$\dot{C}_1(t) = \int_0^t dt' K_1(t-t') C_1(t') + \frac{1}{\sqrt{2}} \int_{-\Delta t}^0 dt' K_1(t-t') C_E(t') \tag{7}$$

with $K_1(t-t') = K(t-t') + K_{AB}(t-t')$, and $K_{AA}(t-t') = K_{BB}(t-t') \equiv K(t-t')$, where

$$K_{AB}(t-t') = -\frac{1}{\hbar\pi\varepsilon_0}\int_0^\infty d\omega \frac{\omega^2}{c^2} e^{-i(\omega-\omega_A)(t-t')} \vec{d}_A \, \text{Im}\overline{\overline{G}}(\vec{r}_A,\vec{r}_B,\omega)\vec{d}_B. \tag{8}$$

In the strong coupling regime, $K(t-t')$ can be approximated as[40]

$$K(t-t') \approx -\frac{1}{2}\Gamma\delta\omega_m e^{-i(\omega_m-\omega_A)(t-t')} e^{-\delta\omega_m|t-t'|}, \tag{9}$$

where $\omega_m$ and $\delta\omega_m$ are the resonance frequency and linewidth of the system, respectively. Similarly,

$$K_1(t-t') \approx -\frac{1}{2}(\Gamma+\Gamma_{AB})\delta\omega_m e^{-i(\omega_m-\omega_A)(t-t')} e^{-\delta\omega_m|t-t'|}. \tag{10}$$

At the resonance frequency, $\Gamma_{AB} \approx \Gamma$ can be realized, there is $\Gamma_1 = \Gamma + \Gamma_{AB} \approx 2\Gamma$, and suppose $\Omega_1 = \sqrt{2}\Omega = 2\sqrt{\Gamma\delta\omega_m}$, after differentiating Eq. (7) we arrive at

$$\ddot{C}_1(t) + [i(\omega_m-\omega_A)+\delta\omega_m]\dot{C}_1(t) + (\frac{\Omega_1}{2})^2 C_1(t) = 0. \tag{11}$$

As $\omega_A = \omega_m$, $\Omega_1 > \delta\omega_m$ and the initial condition $C_1(0) = 0$, the solution of Eq. (11) is

$$C_1(t) = \frac{\sqrt{2}\Omega^2(\delta\omega_m e^{-\delta\omega_m \Delta t/2}-\Omega)}{(\Omega^2+\delta\omega_m^2)\sqrt{\Omega_1^2-\delta\omega_m^2}} e^{-\delta\omega_m(t+\Delta t)/2} \sin\frac{\sqrt{\Omega_1^2-\delta\omega_m^2}}{2}t. \tag{12}$$

Following the same procedure with $C_1(t)$, we can obtain

$$C_2(t) = \frac{\sqrt{2}\Omega^2(\delta\omega_m e^{-\delta\omega_m \Delta t/2}-\Omega)}{(\Omega^2+\delta\omega_m^2)\sqrt{\Omega_2^2-\delta\omega_m^2}} e^{-\delta\omega_m(t+\Delta t)/2} \sin\frac{\sqrt{\Omega_2^2-\delta\omega_m^2}}{2}t, \tag{13}$$

where $K_2(t-t') = K(t-t') - K_{AB}(t-t')$, $\Gamma_{AB} \approx -\Gamma$ and $\Gamma_2 = \Gamma - \Gamma_{AB} \approx 2\Gamma$ are taken. Here $\Omega_2 = \Omega_1 = \sqrt{2}\Omega$ and the line-width $\delta\omega_m$ is determined by the imaginary part of the eigen-frequency[41], which can be obtained by solving the eigenvalue of the system (see supplementary material for the detailed process). Having obtained $\delta\omega_m$ and $\Gamma$, the Rabi frequency $\Omega$ can also be calculated. In general, when $\Omega \sim 10 \cdots 100\delta\omega_m$, the strong coupling can be realized[38]. Our calculated results show that the strong coupling condition can be reached in above systems when the separation distance is small, like d=1nm. Then, from Eqs. (12) and (13) combining with Eqs. (4)-(6), we can calculate the genuine entanglement for two-qubit system with $\Gamma_{AB} \approx \pm\Gamma$.

Figure 1 shows the calculated results when two-level QEs located at the hotspots of a linearly arranged silver nanosphere trimer with various separation distances. The radii of silver spheres are taken as R=10nm. For the dielectric functions of Ag, the Johnson's data were adopted (the absorption loss is included)[42]. The parameters of the molecular dipole of the QEs are taken according to Ref. 43:

$|\vec{d}_A| = |e|r_0$ and $r_0 = 10$ Å. Figure 1 (a), (b) and (c) correspond to the genuine entanglement with d=4nm, d=2nm and d=1nm at different frequencies, respectively. Comparing them, we find that both the amplitude and duration of quantum beats for the genuine entanglement increase with the decrease of the separation distance d. For example, as d=4nm, the maximum value of $E_G^{(2)}$ is less than 0.1, it reaches 0.63 at d=1nm. The duration of quantum beats for such a case can reach two orders longer than that for the entanglement in a photonic cavity (CQED)[32]. These results are for the genuine entanglement. In fact, another popular method to scale two-qubit entanglement has been given in Refs. 44 and 45, where concurrence of two qubits need be calculated. Comparing the calculated results from two kinds of method, we find that they are identical.

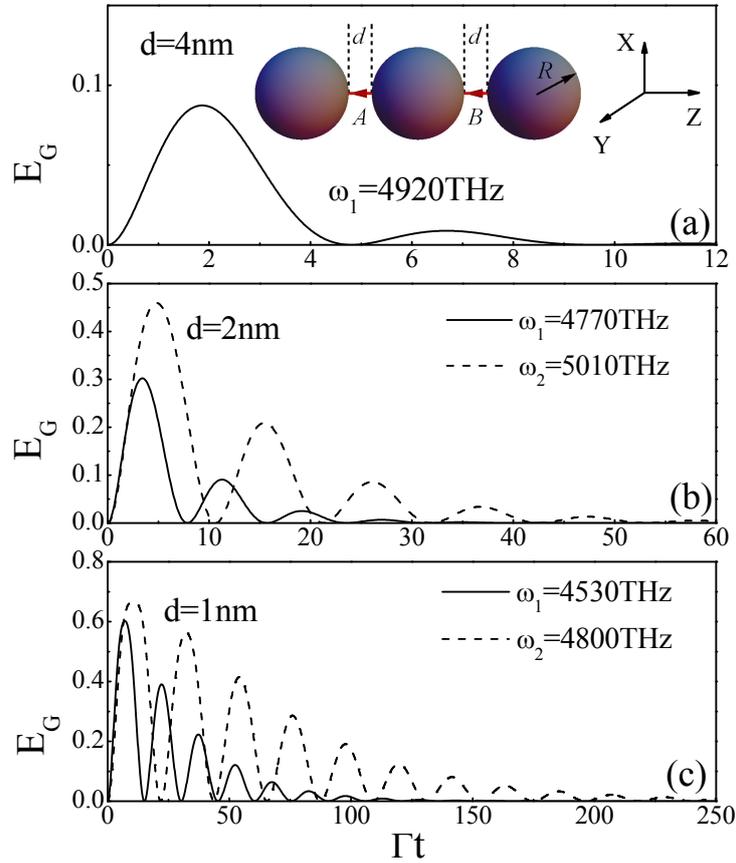

FIG. 1. (Color online) Genuine entanglement of two QEs inserted in the gaps of a linearly arranged silver nanosphere trimer (shown in inset) with various separation distances and frequencies: (a) d=4nm and $\omega_1$=4920THz ; (b) d=2nm, $\omega_1$=4770THz (solid), $\omega_2$=5010THz (dashed); (c) d=1nm, $\omega_1$=4530THz (solid), $\omega_2$=4800THz (dashed). The radii of nanospheres are R=10nm, and the orientations of electric dipole moments for two atoms are both along the axis of the trimer.

Another feature is that the entanglement for two-level QEs can appear at different frequencies simultaneously with the decrease of d. As d=4nm, the entanglement only appears at one frequency $\omega_1 = 4920 THz$ (Fig. 1(a)), it appears at $\omega_1 = 4770 THz$ (solid line in Fig. 1(b)) and $\omega_2 = 5010 THz$ (dashed line in Fig. 1(b)) simultaneously for the case with d=2nm, at $\omega_1 = 4530 THz$ (solid line in Fig. 1(c)) and $\omega_1 = 4800 THz$ (dashed line in Fig. 1(c)) for the case with d=1nm.

In order to disclose the physical origin of the above phenomenon, in Fig. 2(a) and (b) we plot the corresponding single QE spontaneous decay rate $\Gamma$ and the interference term $\Gamma_{AB}/\Gamma$ between QEs as a function of frequency at various gaps. The solid, dashed and dotted line correspond to the case with d=1nm, 2nm and 4nm, respectively. The value of decay rate $\Gamma$ has a whole decrease with the increase of d. It is shown clearly that there exist three enhanced peaks (marked by (1), (2) and (3) for d=1nm) for each case. In these peaks, the position of peak 3 is not sensitive to the gaps, it also corresponds to the calculated result without the hotspots (red dotted line), that is, two QEs are arranged around a metallic sphere as shown in the inset of Fig. 2(a). It is caused by the localized surface plasmon resonance of single metallic sphere. The other two peaks (peak (1) and (2)) are very sensitive to the gaps. With the decrease of the gaps, the peaks shift to longer wavelength (redshift), which correspond to the case of plasmon coupling resonances. This can be seen more clearly from Fig. 3.

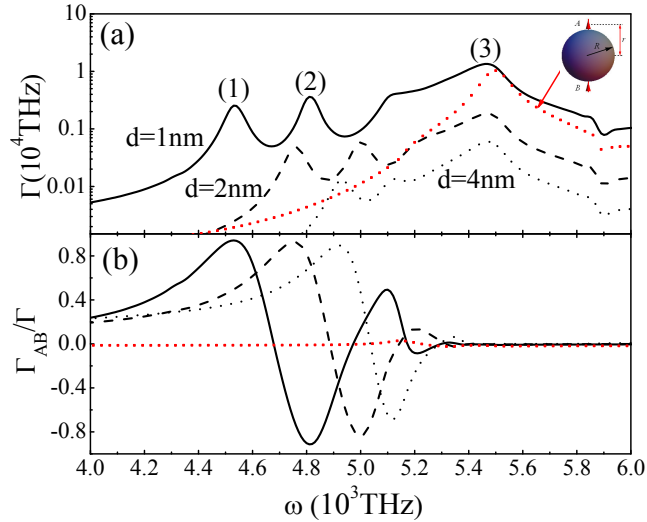

FIG. 2. (Color online) (a) Spontaneous decay rate $\Gamma$ and (b) interference term $\Gamma_{AB}/\Gamma$ of atoms in the structures described in Fig. 1. The solid line, dashed line and dotted line correspond to the cases with d=1nm, 2nm and 4nm, respectively. The other parameters are identical with those in Fig. 1. Red dotted line represents the corresponding result for a single sphere and two QEs without hotspots as shown in inset.

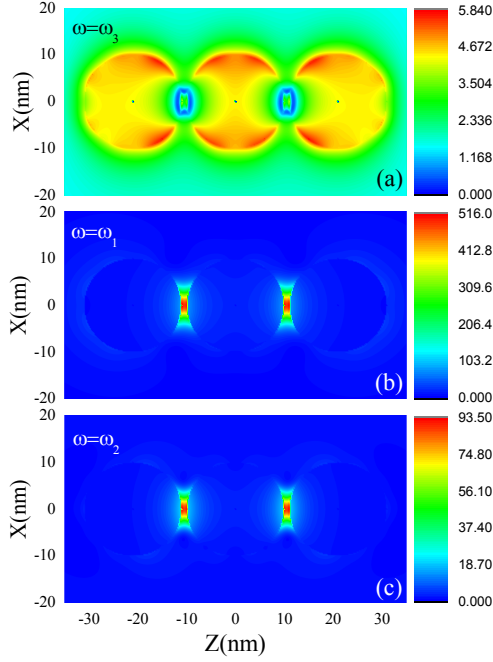

**FIG. 3. (Color online) Electric field amplitude pattern of a linearly arranged silver nanosphere trimer with d=1nm at $\omega_3 = 5470\,\text{THz}$ (a), $\omega_1 = 4530\,\text{THz}$ (b) and $\omega_2 = 4800\,\text{THz}$ (c) for the normal incidence, which correspond to three peaks in Fig. 2(a).**

Figure 3 shows the comparison of local electric filed intensity in XZ plane at three different wavelengths (corresponding to three peaks) for the case with d=1nm. Figure 3(a) corresponds to the result of the single scattering resonance ($\omega_3 = 5470\,\text{THz}$), localized surface plasmon excitation is around the single sphere and the fields in the gaps are not strong. In such a case, the interference term $\Gamma_{AB}/\Gamma$ (red dotted line in Fig. 2(b)) between QEs is almost zero although the single QE spontaneous decay rate is very big. Thus, the entanglement between two QEs cannot be produced. In contrast, the electric fields mainly focus in the gap regions for the coupling resonances as shown in Fig. 3 (a) and (b) (corresponding to peaks (1) and (2) in Fig. 2(a)). The coupling resonance causes big interference term between QEs, which directly leads to the generation of strong entanglement (>0.5) between QEs.

## IV. THREE-QUBIT ENTANGLEMENT

If we consider three two-level QEs (A, B and C) located in the gaps of the metallic sphere cluster as shown in the inset of Fig. 4(a), three-qubit entanglement can be realized. Here, the orientations of the electric-dipole moments are all along the axis of any two spheres. The new states $|i\rangle$ ($i = 1, 2, 3$) are taken as $|1\rangle = \frac{1}{\sqrt{3}}(|A\rangle+|B\rangle+|C\rangle)$, $|2\rangle = \frac{1}{\sqrt{6}}(2|A\rangle-|B\rangle-|C\rangle)$ and $|3\rangle = \frac{1}{\sqrt{6}}(2|B\rangle-|A\rangle-|C\rangle)$. The decay rates of the three states are $\Gamma_1 = \Gamma + 2\Gamma_{AB}$ and $\Gamma_{2(3)} = \Gamma - \Gamma_{AB}$. At $\omega_1 = 4460\,\text{THz}$ corresponding to the peak (1) marked in Fig. 4(b), $\Gamma_{AB} \approx \Gamma$, we obtain $\Gamma_1 \approx 3\Gamma$ and $\Gamma_{2(3)} \approx 0$, the density operator possesses the same form as Eq. (4) but $i$ is replaced by 1, where the probability amplitude $C_1(t)$ can be solved similar to the case for the two-QE regime. When $\omega_A = \omega_m$, $3\Omega^2 > \delta\omega_m^2$, and with the initial condition $C_1(0) = 0$, the solution is

$$C_1(t) = \frac{\sqrt{3}\Omega^2(\delta\omega_m e^{-\delta\omega_m \Delta t/2} - \Omega)}{(\Omega^2 + \delta\omega_m^2)\sqrt{3\Omega^2 - \delta\omega_m^2}} e^{-\delta\omega_m(t+\Delta t)/2} \sin\frac{\sqrt{3\Omega^2 - \delta\omega_m^2}}{2}t . \tag{14}$$

At $\omega_2 = 4710\,\text{THz}$ for the peak 2 in Fig. 4(b), $\Gamma_{AB} \approx -\frac{1}{2}\Gamma$, we have $\Gamma_1 = 0$ and $\Gamma_{2(3)} = \frac{3}{2}\Gamma$, the density operator can also be expressed by Eq. (4) with $i = 2(3)$. Similar to the procedure in solving $C_1(t)$, we can obtain

$$C_{2(3)}(t) = \frac{\sqrt{6}}{2} e^{-(\delta\omega_m t+\Delta t)/2} \frac{\Omega^2(\delta\omega_m e^{-\delta\omega_m \Delta t/2} - \Omega)}{(\Omega^2 + \delta\omega_m^2)\sqrt{\frac{3}{2}\Omega^2 - \delta\omega_m^2}} \sin\frac{\sqrt{\frac{3}{2}\Omega^2 - \delta\omega_m^2}}{2}t . \tag{15}$$

From Eqs. (14) and (15) combining with Eqs. (4)-(6), the genuine three-qubit entanglement can be obtained. The calculated results are shown in Fig. 4(a). The corresponding results for the single QE spontaneous decay rate $\Gamma$ and the interference term $\Gamma_{AB}/\Gamma$ between QEs are given in Fig. 4(b) and (c), respectively. When d=1nm, $\Gamma_{AB} \approx \Gamma$ and $\Gamma_{AB} \approx -\frac{1}{2}\Gamma$ can be reached for the strong coupling condition. In such a case, strong three-qubit entanglement (>0.5) can be observed clearly. However, with the increase of d, for example, when d=2nm or 4nm, the single QE decay rate $\Gamma$ becomes very small, and $\Gamma_{AB}/\Gamma$ is also less than 0.8 for all frequencies, in these cases strong three-qubit entanglements cannot be found. Compared with two-qubit entanglement, the realization of three-qubit entanglement needs stronger resonance coupling, that is, the smaller gaps. With the decrease of d, we have found more peaks appear, such as peaks 3 and 4 marked in Fig. 4(b), which are caused by the collective resonance of three spheres. However considerable entanglement at the frequencies corresponding to these peaks cannot be produced because $\Gamma_{AB}/\Gamma$ is very small as shown in Fig. 4(c). Therefore, to achieve strong multi-qubit entanglement, two conditions, large $\Gamma$ and $\Gamma_{AB}/\Gamma$, must be satisfied simultaneously.

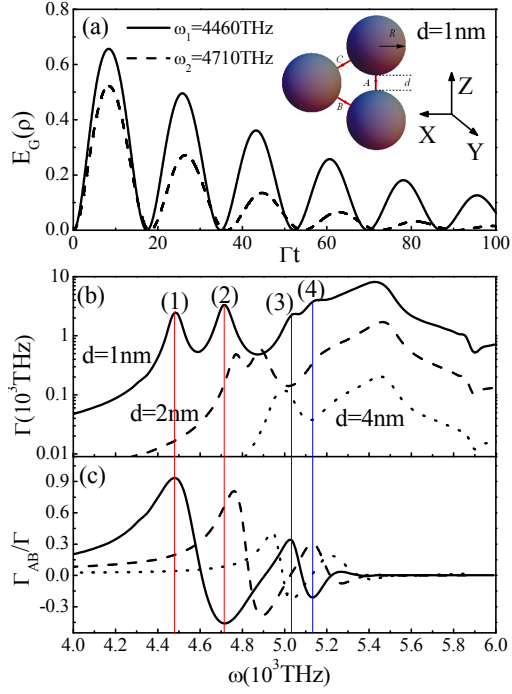

FIG. 4. (Color online) (a) Genuine entanglements of three QEs inserted in the hotspots of three spheres arranged in a triangle configuration showed in the inset at $\omega_1 = 4460\text{THz}$ and $\omega_2 = 4710\text{THz}$. The corresponding spontaneous decay rate $\Gamma$ (b) and interference term $\Gamma_{AB}/\Gamma$ (c) of QEs with d=1nm (solid), d=2nm (dashed) and d=4nm (dotted), the other parameters are taken identical with those in Fig. 1.

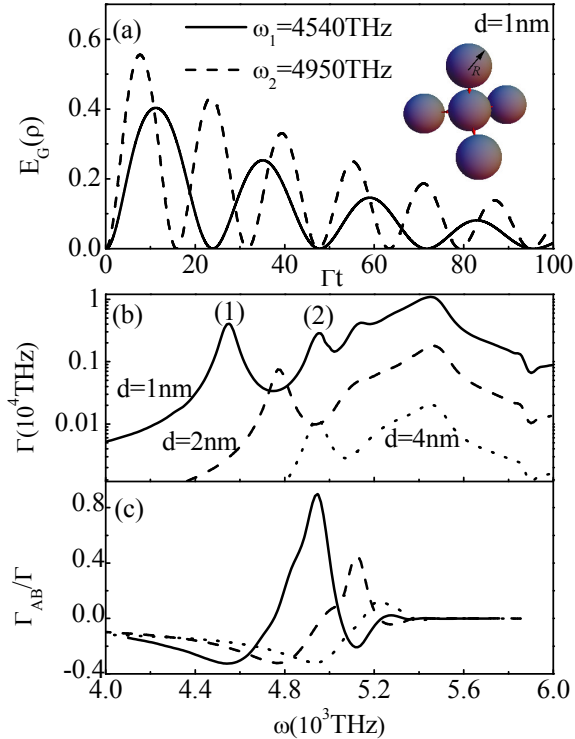

**FIG. 5.** (Color online) (a) Genuine entanglements of three QEs inserted in the hotspots of five spheres cluster as showed in the inset at $\omega_1 = 4540 \text{THz}$ and $\omega_2 = 4950 \text{THz}$. The corresponding spontaneous decay rate $\Gamma$ (b) and interference term $\Gamma_{AB}/\Gamma$ (c) of QEs with d=1nm (solid), d=2nm (dashed) and d=4nm (dotted), the other parameters are taken identical with those in Fig. 1.

## V. FOUR-QUBIT ENTANGLEMENT

Similar to the two-qubit and three-qubit cases, we can also realize four-qubit entanglement. We consider four two-level QEs (A, B, C and D) located in the gaps of the metallic sphere cluster as shown in the inset of Fig. 5(a), to ensure the symmetry, the four spheres are in the vertex of a tetrahedron, with another sphere in the center. The orientations of the electric-dipole moments are all along the axis of any two spheres. The new states $|i\rangle$ ($i = 1, 2, 3, 4$) are taken as $|1\rangle = \frac{1}{\sqrt{4}}(|A\rangle+|B\rangle+|C\rangle+|D\rangle)$, $|2\rangle = \frac{1}{\sqrt{12}}(3|A\rangle-|B\rangle-|C\rangle-|D\rangle)$, $|3\rangle = \frac{1}{\sqrt{12}}(3|B\rangle-|A\rangle-|C\rangle-|D\rangle)$ and $|4\rangle = \frac{1}{\sqrt{12}}(3|C\rangle-|A\rangle-|B\rangle-|D\rangle)$. The decay rates of the four states satisfy $\Gamma_1 = \Gamma + 3\Gamma_{AB}$ and $\Gamma_{2(3,4)} = \Gamma - \Gamma_{AB}$. At $\omega_1 = 4540 \text{THz}$ corresponding to the peak (1) in Fig. 5(b) for d=1nm, $\Gamma_{AB} \approx -\frac{1}{3}\Gamma$, we obtain $\Gamma_1 \approx 0$ and $\Gamma_{2(3,4)} \approx \frac{4}{3}\Gamma$, the density operator is expressed by Eq. (4) with $i = 2(3,4)$, in which $C_{2(3,4)}(t)$ are given as

$$C_{2(3,4)}(t) = \sqrt{\frac{4}{3}} \frac{\Omega^2(\delta\omega_m e^{-\delta\omega_m \Delta t/2} - \Omega)}{(\Omega^2 + \delta\omega_m^2)\sqrt{\frac{4}{3}\Omega^2 - \delta\omega_m^2}} e^{-\delta\omega_m(t+\Delta t)/2} \sin\frac{\sqrt{\frac{4}{3}\Omega^2 - \delta\omega_m^2}}{2}t . \qquad (16)$$

When the frequency is taken corresponding to the peak 2 in Fig. 5(b), $\Gamma_{AB} \approx \Gamma$, we have $\Gamma_1 \approx 4\Gamma$ and $\Gamma_{2(3,4)} \approx 0$, the density operator can be written as Eq. (4) with $i = 1$, and $C_1(t)$ can be solved as

$$C_1(t) = \frac{2\Omega^2(\delta\omega_m e^{-\delta\omega_m \Delta t/2} - \Omega)}{(\Omega^2 + \delta\omega_m^2)\sqrt{4\Omega^2 - \delta\omega_m^2}} e^{-\delta\omega_m(t+\Delta t)/2} \sin\frac{\sqrt{4\Omega^2 - \delta\omega_m^2}}{2}t . \qquad (17)$$

Figure 5(a) shows the calculated results for the genuine four-qubit entanglement at d=1nm from Eqs. (16), (17) and (4)-(6). The corresponding results for the single QE spontaneous decay rate $\Gamma$ and the interference term $\Gamma_{AB}/\Gamma$ between QEs are given in Fig. 5(b) and (c), respectively. Similar to the case of three QEs, when the gaps become very small such as d=1nm, the conditions: $\Gamma_{AB} \approx \Gamma$ and

$\Gamma_{AB} \approx -\frac{1}{3}\Gamma$ for strong four-qubit entanglement are satisfied. With the increase of d, both $\Gamma$ and $\Gamma_{AB}/\Gamma$ decrease rapidly, the corresponding entanglement is very small.

The above discussions only focus on two-, three- and four-qubit cases. In fact, our theory is suitable for designing any multi-qubit entanglement based on plasmonic hotspots. In addition, in the previous studies on the two-qubit entanglement mediated by one-dimensional plasmonic waveguides[32, 34], it has been pointed out that a continuous laser pumping can be used to have a stationary state with a high degree of entanglement. In the present cases, similar method can also be used. Such a multi-qubit entanglement exhibits many advantages in comparing with other schemes for achieving entanglement. For example, it not only possesses longer duration of quantum beats, it is easy to be realized. Recently, some clusters of gold nanospheres, i.e. the dimers/trimers/tetramers, were fabricated successfully by using cysteine chiral molecules as linkers at the hotspots[19]. We expect our design for the multi-qubit entanglement can be realized and the phenomenon can be observed experimentally in the future.

## VI. CONCLUSIONS

In summary, we have presented a theoretical method to study strong coupling between an ensemble of QEs and surface plasmons excited by the nanoparticle cluster using the rigorous first-principles electromagnetic Green's tensor technique. The method is suitable for designing any multi-qubit entanglement for two level QEs, although our discussions focus on two-, three- and four-qubit cases. Such a method for achieving multi-qubit entanglements exhibits many advantages in comparing with other schemes. For example, the multi-qubit entanglement for two-level QEs can be produced at different frequencies simultaneously, when they locate in hot spots of metallic nanoparticle clusters. The duration of quantum beats for such an entanglement can reach two orders longer than that for the entanglement in a photonic cavity. The phenomena originates from collective coupling resonance excitation of the cluster. In contrast to some previous investigations, we have also found that the entanglement between two QEs cannot be produced at the resonance excitation of the single scattering although the single QE spontaneous decay rate is very big. Potential applications of the present phenomena to quantum-information processing are anticipated.

**ACKNOWLEDGMENTS**

This work was supported by the National Natural Science Foundation of China (11274042 and

61421001).

# References

[1] C. H. Bennett, G. Brassard, C. Crepeau, R. Jozsa, A. Peres, and W. K. Wootters, Phys. Rev. Lett. **70**, 1895 (1993).

[2] C. A. Fuchs, N. Gisin, R. B. Griffiths, C-S. Niu, and A. Peres, Phys. Rev. A **56**, 1163 (1997).

[3] C. H. Bennett and S. J. Wiesner, Phys. Rev. Lett. **69**, 2881 (1992).

[4] D. P. DiVincenzo, Science **270**, 255 (1995).

[5] T. D. Ladd, F. Jelezko, R. Laflamme, Y. Nakamura, C. Monroe and J. L. O'Brien, Nature **464**, 45 (2010).

[6] H. J. Kimble, Nature **453**, 1023 (2008).

[7] N. Gisin, G. Ribordy, W. Tittel and H. Zbinden, Rev. Mod. Phys. **74**, 145 (2002).

[8] V. Giovannetti, S. Lloyd and L. Maccone, Phys. Rev. Lett. **96**, 010401 (2006).

[9] R. Horodecki, P. Horodecki, M. Horodecki and K. Horodecki, Rev. Mod. Phys. **81**, 865 (2009).

[10] T. Monz, et.al., Phys. Rev. Lett. **106**, 130506 (2011).

[11] J. T. Barreiro1, J-D Bancal, P. Schindler, D. Nigg, M. Hennrich, T. Monz, N. Gisin and R. Blatt, Nat. Phys. **9**, 559 (2013).

[12] M. Huber and R. Sengupta, Phys. Rev. Lett. **113**, 100501 (2014).

[13] M. Chen, N. C. Menicucci, and O. Pfister, Phys. Rev. Lett. **112**, 120505 (2014).

[14] Yun-Feng Huang, et.al., Nature Communications **2**, 546 (2011).

[15] Xing-Can Yao, et, al., Nature Photonics **6**, 225 (2012).

[16] Jian-Wei Pan, et.al., Rev. Mod. Phys. **84**, 777 (2012).

[17] M. I. Stockman, Phys. Today **64**, 39 (2011).

[18] J. A. Schuller, E. S. Barnard, W. Cai, Y. Jun, J. S. White, M. L. Brongersma, Nat. Mater. **9**, 193 (2010).

[19] Rong-Yao Wang, et. al., The Journal of Physical Chemistry C **118**, 9690 (2014).

[20] S. Savasta, R. Saija, A. Ridolfo, O. D. Stefano, P. Denti, and F. Borghese, ACS Nano **11**, 6369 (2010).

[21] A. Manjavacas, F. J. García de Abajo, and P. Nordlander, Nano Lett. **11**, 2318 (2011).

[22] A. V. Akimov, A. Mukherjee, C. L. Yu, D. E. Chang, A. S. Zibrov, P. R. Hemmer, H. Park, and M. D. Lukin, Nature **450**, 402 (2007).

[23] G. Y. Chen, Y. N. Chen, and D. S. Chuu, Opt. Lett. **33**, 2212 (2008).

[24]D. Dzsotjan, A. S. Sørensen, and M. Fleischhauer, Phys. Rev. B **82**, 075427 (2010).

[25]A. Huck, S. Kumar, A. Shakoor, and U. L. Andersen, Phys. Rev. Lett. **106**, 096801 (2011).

[26]T. Schwartz, J. A. Hutchison, C. Genet, and T.W. Ebbesen, Phys. Rev. Lett. **106**, 196405 (2011).

[27]S. Aberra-Guebrou, et. al., Phys. Rev. Lett. **108**, 066401 (2012).

[28]A. Salomon, R. J. Gordon, Y. Prior, T. Seideman, and M. Sukharev, Phys. Rev. Lett. **109**, 073002 (2012).

[29]K. Słowik, R. Filter, J. Straubel, F. Lederer, and C. Rockstuhl, Phys. Rev. B **88**, 195414 (2013).

[30]A. González-Tudela, P. A. Huidobro, L. Martín-Moreno, C. Tejedor, and F. J. García-Vidal, Phys. Rev. Lett. **110**, 126801 (2013).

[31]G. Y. Chen, N. Lambert, C. H. Chou, Y. N. Chen, and F. Nori, Phys. Rev. B **84**, 045310 (2011).

[32]A. Gonzalez-Tudela, D. Martin-Cano, E. Moreno, L. Martin-Moreno, C. Tejedor, and F. J. Garcia-Vidal, Phys. Rev. Lett. **106**, 020501 (2011).

[33]D. Martin-Cano, A. Gonzalez-Tudela, L. Martin-Moreno, F. J. Garcia-Vidal, C. Tejedor, and E. Moreno, Phys. Rev. B **84**, 235306 (2011).

[34]H. Zheng and H. U. Baranger, Phys. Rev. Lett. **110**, 113601 (2013).

[35]A. González-Tudela1 and D. Porras, Phys. Rev. Lett. **110**, 080502 (2013).

[36]J. Yang, G. W. Lin, Y. P. Niu, and S. Q. Gong, Opt. Express, **21**, 15618 (2013).

[37]J. Ren, J. Yuan, and X. D. Zhang, J. Opt. Soc. Am. B. **31**, 2, 229 (2014).

[38]H. T. Dung, S. Scheel, D.-G. Welsch, and L. Knöll, J. Opt. B **4**, S169–S175 (2002).

[39]T. R. de Oliveira, G. Rigolin, and M. C. de Oliveira, Phys. Rev. A **73**, 010305 (R) (2006); erratum: Phys. Rev. A **75**, 039901 (E) (2007).

[40]H. T. Dung, L. Knöll and D.-G. Welsch, Phys. Rev. A **62**, 053804 (2000).

[41]H. T. Dung, L. Knöll and D.-G. Welsch, Phys. Rev. A **64**, 013804 (2001).

[42]P. B. Johnson and R. W. Christy, Phys. Rev. B **6**, 4370 (1972).

[43]H. Zhang and A. O. Govorov, Phys. Rev. B **87**, 075410 (2013).

[44]S. Hill and W. K. Wootters, Phys. Rev. Lett. **78**, 5022 (1997).

[45]W. K. Wootters, Phys. Rev. Lett. **80**, 2245 (1998).